\documentclass[aps,superscriptaddress,pre,10pt,twocolumn]{revtex4-1}
\usepackage{graphicx}
\usepackage{epstopdf}
\usepackage{amsmath}
\usepackage{amssymb}
\usepackage[utf8]{inputenc}
\usepackage{graphicx,float,hyperref}
\usepackage{natbib}
\usepackage{graphicx,latexsym} 

\begin{document}

\preprint{APS/123-QED}

\title{Unified  trade-off optimization of a three-level quantum refrigerator}

\author{Kirandeep Kaur}
\email{kirandeep@iisermohali.ac.in}
\affiliation{Department of Physical Sciences, Indian Institute of Science Education and Research, Sector 81, S. A. S. Nagar, Manauli PO 140306, Punjab, India}
\author{Varinder Singh}
\email{vsingh@ku.edu.tr}
\affiliation{Department of Physics, Ko\c{c} University, 34450 Sar\i{}yer, Istanbul Turkey}
\author{Jatin Ghai}
\email{jghai98@gmail.com}
\affiliation{Department of Physics, Punjabi University, Patiala, Punjab, India}
\author{Satyajit Jena}
\email{sjena@iisermohali.ac.in}
\affiliation{Department of Physical Sciences, Indian Institute of Science Education and Research, Sector 81, S. A. S. Nagar, Manauli PO 140306, Punjab, India}
\author{\"{O}zg\"{u}r E. M\"{u}stecapl{\i}o\u{g}lu}
\email{omustecap@ku.edu.tr}
\affiliation{Department of Physics, Ko\c{c} University, 34450 Sar\i{}yer, Istanbul Turkey}
%

\begin{abstract}
We study the optimal performance of a three-level quantum refrigerator using a trade-off objective
function, $\Omega$ function, which represents a compromise between the energy benefits and
the energy losses of a thermal device. First, we optimize the performance of our refrigerator by employing a two-parameter 
optimization scheme and show that the first two-terms in the series expansion of the obtained coefficient of performance (COP) match with those of some classical models of refrigerator. Then, in the high-temperature limit, optimizing with respect to one parameter while constraining the other one, we obtain the lower and upper bounds on the COP for both strong as well as weak (intermediate) matter-field coupling conditions. In the strong matter-field coupling regime, the obtained bounds on the COP exactly match with the bounds already known for some models
of classical refrigerators. Further for weak matter-field coupling, we derive some new bounds on the
the COP of the refrigerator which lie beyond the range covered by bounds obtained for strong
matter-field coupling. Finally, in the parameter regime where both cooling power and $\Omega$ function can be maximized, we compare
the cooling power of the quantum refrigerator at maximum $\Omega$ function with the maximum cooling power.
\end{abstract}

\maketitle


\section{\label{sec:level1}Introduction}

The first theoretical study of a heat engine, operating between two thermal reservoirs at temperature $T_c$ and $T_h $ ($ T_h > T_c $), was carried out by Carnot back in 1824. The abstract Carnot cycle operates at  Carnot efficiency, $ \eta_C = 1 - T_c/T_h$, which serves as theoretical upper bound on the efficiency of all classical macroscopic heat engines. On operating heat cycle in a reverse order, it turns into a refrigerator and the corresponding performance measure  is known as coefficient of performance (COP), $\zeta_C = T_c/(T_h - T_c)$. The practical implications of Carnot efficiency are limited as it can be obtained only in reversible process which is infinitesimally slow, thereby producing vanishing power output, which makes it quite impractical. The search for realistic operational regime of heat engines operating at finite power in finite time gave rise to the development of finite time thermodynamics (FTT) \cite{Berry1984,Salamon2001,Andresen2011}. FTT establishes the best mode of operation of heat engines by conveniently modelling the constraints arising due to the sources of irreversibilities, finite time etc., and then optimizing a suitable objective function with respect to the system parameters. The freedom in choosing the objective function lead the researchers to look for variety of criteria considering thermodynamic sustainability, environmental and economic aspects \citep{ABrown1991}. For instance, Yvon \cite{Yvon} and Novikov \cite{Novikov1958} derived the expression for the efficiency at maximum power of nuclear power plants in mid 1950s. This expression was rederived by Curzon and Ahlborn (CA) \cite{CA1975}  in 1975 for an endoreversible heat engine, obeying Newtonian heat transfer law between the reservoirs and the working material, by using the 
assumption of instantaneos adiabats \cite{Rubin1979A,Rubin1979B}, and is  given by $\eta_{CA} = 1-\sqrt{1-\eta_C}$.  It is a remarkable result as it is independent of system parameters and is in good agreement with the efficiency of actual thermal power plants \cite{CA1975}. Further, Esposito \textit{et al}. obtained the same expression for the 
efficiency at maximum power for the optimization of a symmetric low-dissipation heat engine \cite{Esposito2010}.  

Many attempts have been made to set a similar model independent benchmark for the optimal COP for refrigerators \cite{YanChen1989,AgarwalMenon,deTomas2012,Apertet2013A}, but it turns out that it is not straight-forward to perform optimization analysis of the refrigerators . For many models of refrigerator, optimization of the cooling power (CP) of the refrigerator \cite{YanChen1989,YanChen1990,deTomas2012,Allahverdyan2010}, which replaces power as the figure of merit, is not an appropriate objective function to optimize \cite{YanChen1989,LutzEPL,Johal2019}.  For instance, optimization of CP with Newton's heat transfer laws for the endoreversible model results in vanishing COP, which does not have any practical significance. It can be numerically optimized to produce finite value of the COP only when one accounts for the time spent on the adiabatic branches of the cycle \cite{AgarwalMenon}. Similarly, 
for the low-dissipation refrigerators,  a generic maximum for CP does not exist. However, by minimizing the input work
first, the CP can be maximized for a given cycle time \cite{Johal2019}.

Working along the lines of FTT, Yan and Chen \cite{YanChen1989} proposed a new criterion, $ \chi = \zeta\dot{Q_c}$ which gives equal emphasis on COP and CP, for the optimization of endoreversible refrigerators. The COP for optimal $\chi$ criterion, using Newtons heat transfer law within endoreversible approximation, is given by
\begin{equation}
    \zeta_{CA} = \sqrt{1 + \zeta_C} - 1 . 
    \label{CA} 
\end{equation}
This expression holds for both classical \cite{YanChen1989,deTomas2012} and quantum \cite{Allahverdyan2010,LutzEPL,VJ2020}  models of refrigerator. 
Taking the optimization analysis of refrigerators one step ahead, Hern\'andez \textit{et al.} \cite{Hernandez2001} proposed a new unified trade-off optimization criterion, $\Omega$-criterion, which is easy to implement both in heat engines and refrigerators, and amenable to analytic results. $\Omega$ function is a trade-off objective function which represents a compromise between the energy benefits and the energy losses. The optimization of $\Omega$ function yields the COP that lies in between the region of maximum COP and the COP at maximum CP. For steady state refrigerators, $\Omega$ function is defined as follows  \cite{Hernandez2001}
\begin{equation}
    \Omega = (2\zeta-\zeta_{\rm max})\dot{Q}_c/\zeta,
    \label{Omega}
\end{equation}
where $\dot{Q}_c$ is cooling power; $\zeta$ and $\zeta_{\rm max}$ are the COP and maximum value of COP for 
the given setup. In this work, we  make a detailed study of the three-level 
refrigerator operating under the conditions of maximum $\Omega$ function (MOF), and obtain analytic expressions
for the COP under various operational conditions. The choice of the model is motivated by its simplicity 
and amenability to analytic results.

The paper is organized as follows. In Sec.~\ref{sec:level2} we discuss the model of three-level quantum refrigerator(SSD). In Sec.~\ref{sec:level3} we obtain the general expression for COP operating at  MOF and find the lower and upper bounds of COP for two optimization schemes in different temperature and coupling regimes. WE compare the CP at MOF with the optimal CP in Sec.~\ref{sec:level4} We conclude in Sec.~\ref{sec:level5}.\\\\

\section{\label{sec:level2}Model of three level laser quantum refrigerator}
\begin{figure} 
\begin{center}
\includegraphics[width=8.6cm]{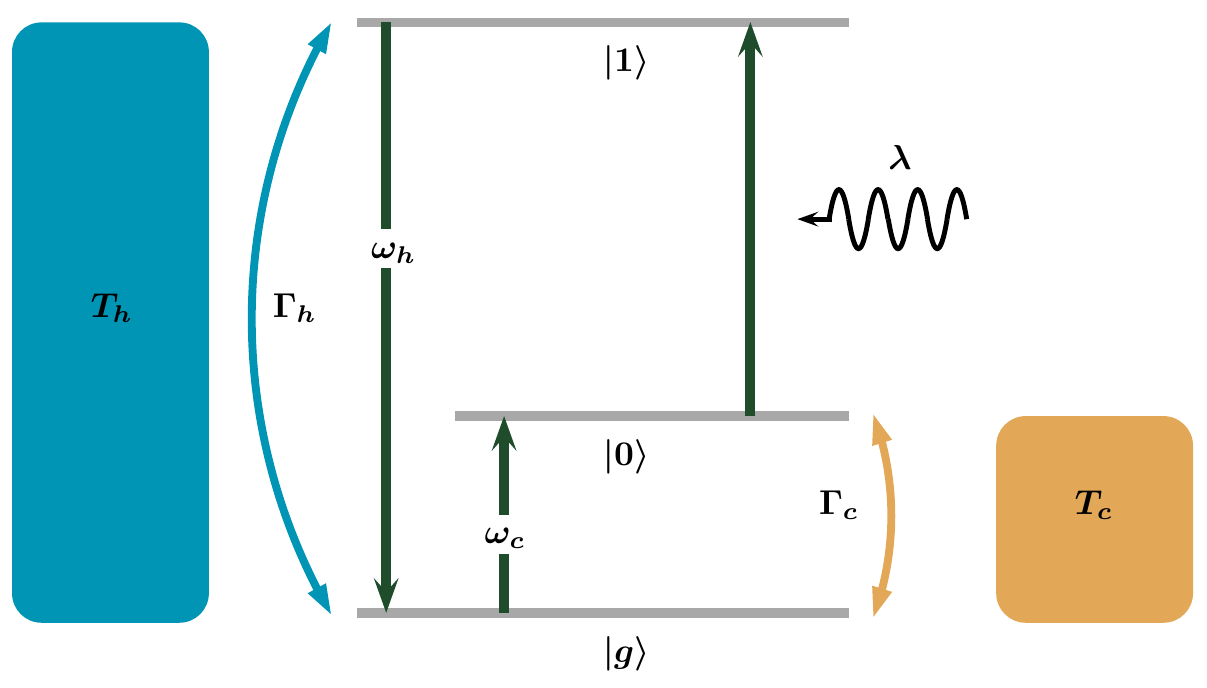}
\end{center}
\caption{  Schematic of three-level refrigerator continuously coupled to two heat reservoirs at temperatures $T_c$ and $T_h$ with coupling constants $\Gamma_c$ and $\Gamma_h$, respectively. A single mode classical field drives the transition between $\vert$0$\rangle$ and $\vert$1$\rangle$, and $\lambda$ represents the strength of matter-field coupling.}
\label{model}
\end{figure}
A three-level heat engine (refrigerator) is one of the simplest model of quantum and has been studied extensively in the literature \cite{Scovil1959B,Geva1994,Geva1996,Geva2002,Scully2001,ScullyAgarwal2003,Linke2005,
BoukobzaTannor2006A,BoukobzaTannor2006B,BoukobzaTannor2007,Scully2011,HarbolaEPL,Harbola2013,Uzdin2015,Harris2016,
KimAgarwal,Cleuren2012,AC2016,Ghosh2017,Ghosh2018,Dorfman2018,TahirOzgur2019,VJ2019,VJ2020,Brandner2020,Linden2010,Levy2012,Alonso2013,Bijay2017,Segal2018,Scarani2019,Mitchison2016,Brunner2015,TahirOzgur2020}. It consists of a three-level system coupled simultaneously to two thermal reservoirs and a single mode classical field  (Fig. 1). 
The system absorbs energy from the cold bath and jumps from level $\vert$g$\rangle$ to level $\vert$0$\rangle$. The power input mechanism, which consists of an external single mode field coupled to the levels $\vert$0$\rangle$ and $\vert$1$\rangle$, excites the transitions between $\vert$0$\rangle$ and $\vert$1$\rangle$. The population in level $\vert$1$\rangle$ then relaxes to level $\vert$g$\rangle$ by rejecting the heat to the hot bath. The Hamiltonian of the system is given by: $H_0=\hbar \sum \omega_k\vert$k$\rangle  \langle$k$\vert$, where the summation is taken over all the three states and $\omega_k$'s represent atomic frequency of the particular state. Under the rotating wave approximation, the interaction with the single mode lasing field of frequency $\omega$ is described by the semiclassical Hamiltonian $V(t)=\hbar\lambda(e^{-i\omega t}\vert$1$\rangle\langle$0$\vert+e^{i\omega t}\vert$0$\rangle\langle$1$\vert)$, where $\lambda$ is the matter-field coupling constant.  The time evolution of the system is governed by the following GKSL master equation \cite{Lindblad,Gorini}:
\begin{equation}
\dot{\rho}=-\frac{i}{\hbar}[H_0+V(t),\rho]+\mathcal{L}_h[\rho]+\mathcal{L}_c[\rho],
\label{rho}
\end{equation}  
where $\mathcal{L}_{h(c)}$ represents the dissipative Lindblad superoperator and describes the interaction of the system with hot(cold) reservoir:
\begin{multline}
\mathcal{L}_h[\rho]=\Gamma_h(n_h+1)(2\vert g \rangle\langle g \vert\rho_{11}-\vert 1 \rangle\langle 1 \vert\rho-\rho\vert 1 \rangle\langle 1 \vert)\\
+\Gamma_hn_h(2\vert 1 \rangle\langle 1 \vert\rho_{gg}-\vert g \rangle\langle g \vert\rho-\rho\vert g \rangle\langle g \vert),
\end{multline}
\begin{multline}
\mathcal{L}_c[\rho]=\Gamma_c(n_c+1)(2\vert g \rangle\langle g \vert\rho_{00}-\vert 0 \rangle\langle 0 \vert\rho-\rho\vert 0 \rangle\langle 0 \vert)\\
+\Gamma_cn_c(2\vert 0 \rangle\langle 0 \vert\rho_{gg}-\vert g \rangle\langle g \vert\rho-\rho\vert g \rangle\langle g \vert).
\end{multline}
Here $\Gamma_{h}$ and $\Gamma_c$ are the Weisskopf-Wigner decay constants, and $n_{h(c)}=1/(exp[\hbar\omega_{h(c)}/k_BT_{h(c)}]-1)$ is the average number of photons in the hot(cold) reservoir satisfying the relations $\omega_c=\omega_0-\omega_g,\omega_h=\omega_1-\omega_g$. 
We can find a rotating frame for this model in which the steady-state density matrix is time-independent\cite{BoukobzaTannor2007}. Defining $\bar{H}=\hbar(\omega_g\vert$g$\rangle\langle$g$\vert-\frac{\omega}{2}\vert$0$\rangle\langle$0$\vert+\frac{\omega}{2}\vert$1$\rangle\langle$1$\vert)$, an arbitrary operator A in this frame is given by $A_R=e^{i\bar{H}t}Ae^{-i\bar{H}t}$. It can be verified that $\mathcal{L}_h[\rho]$ and $\mathcal{L}_c[\rho]$ remain unchanged under this transformation. Time evolution of the density matrix in this frame can be written as:
\begin{equation}
\dot{\rho_R}=-\frac{i}{\hbar}[H_0-\bar{H}+V_R,\rho_R]+\mathcal{L}_h[\rho_R]+\mathcal{L}_c[\rho_R],
\end{equation}
where $V_R=\hbar\lambda(\vert$1$\rangle\langle$0$\vert+\vert$0$\rangle\langle$1$\vert)$.
Following the work of Boukobza and Tannor \cite{BoukobzaTannor2007,BoukobzaTannor2006A,BoukobzaTannor2006B}, in the following we define the input power and CP of the refrigerator foe a weak system-bath coupling \cite{Alicki1979}:
\begin{equation}
P=\frac{i}{\hbar}Tr([H_0,V_R]\rho_R),
\end{equation}
\begin{equation}
\dot{Q_c}=Tr(\mathcal{L}_c[\rho_R]H_0).
\end{equation} 
Calculating these traces (see Appendix A), the input power and heat flux can be written as:
\begin{equation}
P=i\hbar\lambda(\omega_h-\omega_c)(\rho_{10}-\rho_{01}),
\label{power}
\end{equation} 
\begin{equation}
\dot{Q_C}=i\hbar\lambda\omega_c(\rho_{10}-\rho_{01}),
\label{cool}
\end{equation}
where $\rho_{10}=\langle$1$\vert\rho_R\vert$0$\rangle$ and $\rho_{01}=\langle$0$\vert\rho_R\vert$1$\rangle$. Then, the COP is given by
\begin{equation}
\zeta=\frac{\dot{Q_c}}{P}=\frac{\omega_c}{\omega_h-\omega_c},
\label{COP}
\end{equation}
which satisfies $\zeta\leq\zeta_C$. Hence $\zeta_{\rm max}=\zeta_C$.    The COP of the SSD refrigerator depends upon $\omega_c$ and $\omega_h$ only. Therefore, we choose them as control parameters to study the performance of SSD refrigerator. 

\section{\label{sec:level3} OPTIMIZATION OF $\Omega$ FUNCTION}

In this section, we optimize $\Omega$ function under various operational regimes and find   the analytic expressions for  
the corresponding COPs. Using Eqs. (9) and (10) in equation (2), we can write $\Omega$ function as
\begin{equation}
\Omega=i\hbar\lambda[(2+\zeta_C)\omega_c-\zeta_C\omega_h](\rho_{10}-\rho_{01}).
\end{equation}
The general expression for $\Omega$ function is obtained in Appendix A and is given by Eq. (A12). It is not possible to find an analytic expression for the optimal COP by optimizing Eq. (A12). Therefore, we study the performance of our refrigerator in  
low- and high-temperature regimes in which it is possible to obtain closed form of the COP. Henceforth, for the calculation
purposes, we set $\hbar=k_B=1$.
\subsection{Optimization in low temperature regime}
We begin our optimization analysis in the low temperature regime by assuming $\omega_{c(h)}\gg T_{c(h)}$ and hence setting $n_{c(h)}=e^{-\omega_{c(h)}/T_{c(h)}}\ll 1$.  Under these conditions, Eq. (A12)  takes the following form:
\begin{equation}
\Omega =\dfrac{2\lambda^{2}\Gamma_c\Gamma_h(n_c-n_h)[(2+\zeta_C)\omega_c-\zeta_C\omega
_h]}{(\Gamma_c+\Gamma_h)(\lambda^{2}+\Gamma_c\Gamma_h)}.
\label{omega low}
\end{equation}
Here we will perform a two-parameter optimization of the $\Omega$ function by setting $\partial\Omega/\partial\omega_c =    \partial\Omega/\partial\omega_h = 0$. Solving the resulting equations, we obtain the optimal values of $\omega_c$ and $\omega_h$ as (see Appendix C)
\begin{equation}
\omega_c^*=[1-(1+\zeta_C)k]T_c,\, \omega_h^*=\frac{(1+\zeta_C)[1-(2+\zeta_C)k]T_c}{\zeta_C},
\label{wc wh}
\end{equation}
where $k=ln[(1+\zeta_C)/(2+\zeta_C)]$. The expression for the COP can be obtained by substituting above expressions for 
$\omega_c^*$ and $\omega_h^*$ in Eq. (11), and is given by
\begin{equation}
\zeta_{SSD}=\frac{1-(1+\zeta_C)k}{1-2(1+\zeta_C)k}\zeta_C.
\label{cop low}
\end{equation}
Note that the expression for the COP of SSD refrigerator, $\zeta_{SSD}^{\Omega}$, depends on the ratio of reservoir temperatures only and does not show any dependence on the system parameters. The expression also holds for the optimization of  Feynman's ratchet and pawl model \cite{VarinderJohal}, a classical heat engine based on the principle of Brownian fluctuations. We are also interested in comparing the behavior of the  SSD model with some classical models of refrigerator. This can be done by observing the series behavior of the respective forms of the COP near equilibrium. The series expansions for the COPs of the SSD model, classical endoreversible (low-dissipation) \cite{YanChen1989,deTomas2013} and minimally nonlinear irreversible (MNI) models \cite{LongLiu2014} are given by following expressions, respectively:
\begin{widetext}
\begin{eqnarray}
\zeta_{\rm SSD}  &= & \frac{1-(1+\zeta_C)k}{1-2(1+\zeta_C)k}\zeta_C = \frac{2\zeta_C}{3}+\frac{1}{18}-\frac{16}{216\zeta_C}+O \left(\frac{1}{\zeta_C^{2}}\right),
\label{cop series}
\\
\zeta_{\rm YC} &=&   \sqrt{1+\zeta_C}-1=\frac{2\zeta_C}{3}+\frac{1}{18}-\frac{17}{216\zeta_C}+O \left(\frac{1}{\zeta_C^{2}}\right),
\label{cop AB series}
\\
\zeta_{\rm MNI} &=& \frac{3+4\zeta_C}{4+6\zeta_C}\zeta_C =\frac{2\zeta_C}{3}+\frac{1}{18}-\frac{8}{216\zeta_C}+O \left(\frac{1}{\zeta_C^{2}}\right).
\label{cop AB series2}
\end{eqnarray}
\end{widetext}
Remarkably, the first two terms of the above equations are same and the model dependent difference appears in the third term only, owing to which  $\zeta_{\rm SSD}$, $\zeta_{\rm YC}$ and $\zeta_{\rm MNI}$ lie very close to each other. In fact, for heat engines obeying tight-coupling condition (no heat leaks) and possessing a certain left-right symmetry in the system, the universality of first two terms has already been proved formally \cite{Broeck2005,Lindenberg2009}. However, such universal behavior is not common for the optimal performance of the refrigerators, and is exclusive to the optimization of 
$\Omega$ function. Such universal behavior  absent in the optimization of $\chi$-criterion (see Sec. 3B).

\subsection{Optimization in the high-temperature limit}
In many models of quantum thermal devices, high-temperature regime (classical regime) is employed to obtain
the analytic results and then the obtained results are compared  with the corresponding classical models \cite{Kosloff1984,Geva1992,VJ2020,Alonso2014B}. In our model,
in order to obtain  model-independent  bounds on the COP, we have to complement high-temperature regime with some 
restrictions on the matter-field coupling constant $\lambda$. First, we will discuss the case in which matter-field 
coupling is very strong as compared to the system bath coupling, i.e., $ \lambda\gg\Gamma_{h,c}$.

\subsubsection{Strong coupling and high-temperature regime}
In the high-temperature limit, $n_h$ and $n_c$ are approximated by: $n_h=T_h/\omega_h \gg1$ and $n_c=T_c/\omega_c\gg1$, respectively.
Further, in  the presence of a strong matter-field coupling ($\lambda\gg\Gamma_{h,c}$),  
the expression for $\Omega$ [Eq. (A12)] reduces to the form: 
\begin{equation}
\Omega =\frac{2\Gamma_h(\tau\omega_h-\omega_c)((2+\zeta_C)\omega_c-\zeta_C\omega_h)}{3(\gamma\omega_c+\tau\omega_h)},
\label{omega high}
\end{equation}
where $\tau=T_c/T_h$ and $\gamma=\Gamma_h/\Gamma_c$. A two-parameter optimization scheme of $\Omega$ function with respect to   $\omega_c$ and $\omega_h$ simultaneously leads to the trivial result, 
$\omega_c=\omega_h=0$. Therefore, the choice we are left with is to optimize the $\Omega$ function with 
respect to one control parameter only while keeping the other one fixed at a constant value.

First we optimize Eq. (\ref{omega high}) with respect to $\omega_c$ (fixed $\omega_h$), and the resulting form of the COP is
given by (see Appendix B)
\begin{equation}
\zeta_{\omega_h}= \frac{\tau\Big(2-\tau-\sqrt{(1+\gamma)(2-\tau)(2+\gamma-\tau)}  \Big)}
{\gamma(\tau-2)-\tau\Big(2-\tau-\sqrt{(1+\gamma)(2-\tau)(2+\gamma-\tau)}  \Big)}, \label{effwc}
\end{equation}
We are interested in the extreme dissipation cases for which $\gamma\rightarrow$ and 
$\gamma\rightarrow\infty$. Since Eq. (\ref{ax1}) is monotonic decreasing function of $\gamma$, taking the limits $\gamma\to\infty$ and
$\gamma\to0$, and writing in terms of $\zeta_C$, we find that the COP lies in the range:
\begin{equation}
\zeta_{\rm YC}\equiv\frac{\zeta_C}{\sqrt{(2+\zeta_C)(1+\zeta_C)}-\zeta_C}\leq\zeta_{\omega_h}\leq\frac{3+2\zeta
_c}{4+3\zeta_C}\zeta_C \equiv\zeta_{+} \label{bound high wc}.
\end{equation}
The lower bound $\zeta_{\rm YC}$ obtained here was first obtained by Yan-Chen (YC) for the ecological optimization of 
a classical endoreversible refrigerator \cite{YanChenEco}. Hence, we name it after them. $\zeta_{\rm YC}$ can also be obtained for the  unified trade-off optimization of symmetric low-dissipation refrigerators \cite{deTomas2013} and Carnot-like refrigerators with non-isothermal heat exchanging processes \cite{YanGuo2012}.  The upper bound $\zeta_{+}$ obtained here
also serves as the upper bound on the COP of the low-dissipation refrigerators \cite{deTomas2013} and minimally nonlinear irreversible refrigerators \cite{LongLiu2014}.

Similarly, when we optimize $\Omega$ function with respect to $\omega_h$ for a fixed $\omega_c$, 
the lower and upper bounds on the COP are given by
 \begin{equation}
\zeta_{-}\equiv\frac{2\zeta_C}{3}\leq\zeta_{\omega_c}\leq\zeta_{\rm YC}.
\label{bound high wh}
\end{equation}
Again, under the extreme dissipation conditions, the lower bound $\zeta_{-}=2\zeta_C/3$  obtained here also serve 
as the  lower  bounds on the COP of above-mentioned classical models  of refrigerators. 

The corresponding COP bounds for the optimization of $\chi$-function of the SSD refrigerator are given by \cite{VJ2020}:
\begin{eqnarray}
\zeta^{\chi}_-\equiv 0 && \,\leq\zeta^{\chi}_{\omega_c} \leq \zeta_{CA}, 
\\
\zeta_{CA}  \leq\zeta^{\chi}_{\omega_h} &&\leq \frac{1}{2}(\sqrt{9+8\zeta_C}-3)\equiv \zeta^{\chi}_+.
\end{eqnarray}
In Fig. 2, we have plotted Eqs. (21)-(24). From the Fig. 2, we can see that   except for very small values of 
$\zeta_C$ ($\zeta_C<1$), the refrigerator operating under 
MOF is more efficient than the refrigerator operating at maximum $\chi$-function.

\subsubsection{High temperature and weak or intermediate-coupling regime}
Besides strong matter-field coupling, we can obtain the   closed form analytic expressions for COP in the intermediate weak matter-field coupling ($\lambda\ll\Gamma_{h,c}$) regime  or intermediate-coupling 
($\lambda^2=\Gamma_h\Gamma_c$) regime. Under the above-said condition of weak or
intermediate matter-field coupling, Eq. (A12) can be approximated by the following equation:

\begin{figure}[ht]
\begin{center}
\includegraphics[width=8.6cm]{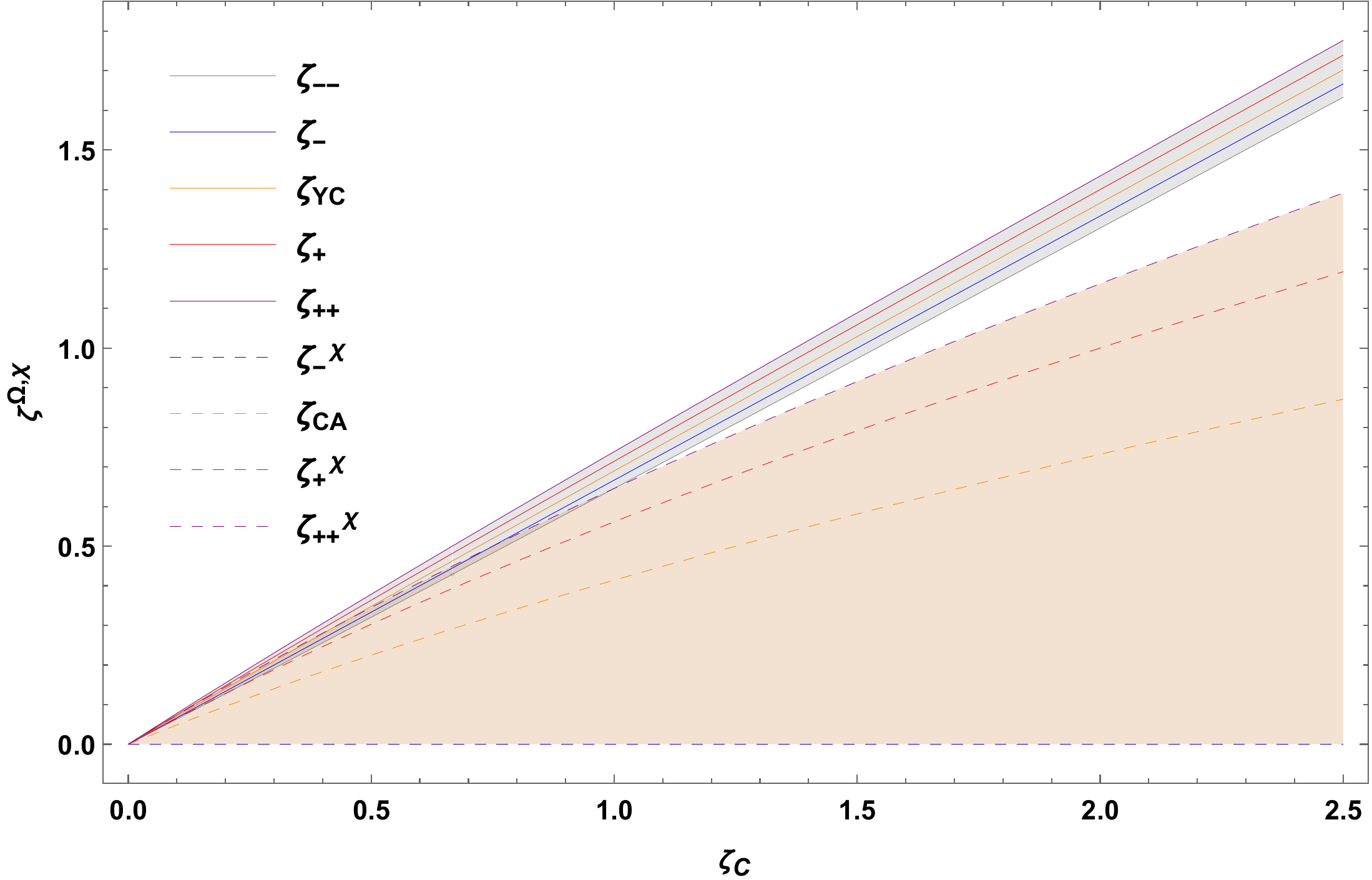}
\end{center}
\caption{COP at MOF versus $\zeta_C$ for a quantum SSD refrigerator in extreme dissipation conditions in high temperature limit and its comparison with COP at maximum $\chi$ function.} \label{COP graph}
\end{figure}  
\begin{equation}
\Omega =\frac{2\lambda^2(n_c-n_h)[(2+\zeta_C)\omega_c-\zeta_C\omega_h]}{3n_hn_{c}^2\Gamma_c+3n_cn_{h}^2\Gamma_h}.
\label{omega high weak}
\end{equation}
\begin{table*}
\caption{Comparison of series expansions for the several forms of COP (efficiency) obtained at  MOF and Maximum $\chi$-function.}
\renewcommand{\arraystretch}{1.5}
\centering
\begin{tabular}{|c|c|c|c|c|}
\hline
COP at  MOF&Efficiency at  MOF&COP at Maximum $\chi$ criteria\\ \hline
$\zeta_{--}=\frac{2}{3}\zeta_C-\frac{1}{18}+\frac{19}{216}\frac{1}{\zeta_C}+O\left(\frac{1}{\zeta_C}\right)^{2}$ 
&
$\eta_{--}=\frac{3}{4}\eta_C-\frac{1}{32}\eta_C^2-\frac{3}{128}\eta_C^3+O(\eta_C)^{4}$
&
$\zeta_{--}^{\chi}=0$ 
\\ \hline
$\zeta^{-}=\frac{2}{3}\zeta_C$ 
&
$\eta_{-}=\frac{3}{4}\eta_C$
&
$\zeta_{-}^{\chi}=0$ 
\\ \hline
$\zeta_{YC}=\frac{2}{3}\zeta_C+\frac{1}{18}-\frac{17}{216}\frac{1}{\zeta_C}+O\left(\frac{1}{\zeta_C}\right)^{2}$
&
$\eta_{}=\frac{3}{4}\eta_C+\frac{1}{32}\eta_C^2+\frac{3}{128}\eta_C^3+O(\eta_C)^{4}$
&
$\zeta_{CA}^{\chi}=\zeta_C^{1/2}-1+\frac{1}{2}\left(\frac{1}{\zeta_C}\right)^\frac{1}{2}
+O\left(\frac{1}{\zeta_C}\right)^\frac{3}{2}$
\\ \hline
$\zeta_{+}   = \frac{2}{3}\zeta_C+\frac{1}{9}-\frac{4}{27}\frac{1}{\zeta_C}+O(\frac{1}{\zeta_C})^{2}$  
&
$\eta_{+}=\frac{3}{4}\eta_C+\frac{2}{32}\eta_C^2+\frac{3}{64}\eta_C^3+O(\eta_C)^{4}$
& 
$\zeta_{+}^{\chi}=\sqrt{2} \zeta_C^{1/2}-\frac{3}{2}+\frac{9}{8\sqrt{2}}\left(\frac{1}{\zeta_C}\right)^{1/2}
+O\left(\frac{1}{\zeta_C}\right)^{3/2}$  
\\ \hline
$\zeta_{++}=\frac{2}{3}\zeta_C+\frac{1}{6}-\frac{5}{24}\frac{1}{\zeta_C}+O\left(\frac{1}{\zeta_C}\right)^{2}$ 
&
$\eta_{++}=\frac{3}{4}\eta_C+\frac{3}{32}\eta_C^2+\frac{9}{128}\eta_C^3+O(\eta_C)^{4}$
& 
$\zeta_{++}^{\chi}=\sqrt{3} \zeta_C^{1/2}-2+\frac{2}{\sqrt{3}}\left(\frac{1}{\zeta_C}\right)^{1/2}
+O\left(\frac{1}{\zeta_C}\right)^{3/2}$\\ \hline
\end{tabular}\label{tab:eng-maaath}
\end{table*}

%
To proceed further, we will use extreme dissipation conditions, i.e., either 
  $\Gamma_c \ll \Gamma_h$ ($\gamma\to\infty$) or
$\Gamma_c\gg\Gamma_h$ ($\gamma\to0$). For the first case ($\Gamma_c\ll\Gamma_h$), we can drop second term in the
denominator of Eq. (\ref{omega high weak}), and the resulting form of $\Omega$ is given as,

\begin{equation}
\Omega_{\gamma\to\infty} =\frac{2\lambda^2(n_c-n_h)[(2+\zeta_C)\omega_c-\zeta_C\omega_h]}{3n_c n_{h}^2\Gamma_h }.
\label{omega inf}
\end{equation} 
Under another extreme dissipation condition $\Gamma_h\ll\Gamma_c$ ($\gamma\to0$), Eq. (23) reads as
\begin{equation}
\Omega_{\gamma\to0} =\frac{2\lambda^2(n_c-n_h)[(2+\zeta_C)\omega_c-\zeta_C\omega_h]}{3n_hn_{c}^2\Gamma_c }.
\label{omega zero}
\end{equation} 
Optimization of Eqs. (\ref{omega inf}) and (\ref{omega zero}) with respect to $\omega_h$ ($\omega_c$ fixed) 
yields the following bounds on the COP:
\begin{equation} 
  \zeta_{--} \equiv \frac{\zeta_C[\zeta_C-3+\sqrt{3+\zeta_C(3+\zeta_C)}]}{3\zeta_C-2}
  \leq      \zeta_{\omega_c}          \leq     \zeta_-.
\end{equation}
The bounds obtained above lie below the parametric region bounded by COP curves given in Eq. (\ref{bound high wh}). It is worthful to mention that these bounds have not been previously obtained for any classical or quantum model of refrigerator.
In the similar manner, optimizing Eqs. (24) and (25) with respect to $\omega_c$ ($\omega_h$ fixed), we obtain  
following bounds on the COP:
\begin{equation}
\zeta_+ 
\leq      \zeta_{\omega_h} \leq
 \frac{\zeta_C[3+\zeta_C+\sqrt{3+\zeta_C(3+\zeta_C)}]}{3(2+\zeta_C)} \equiv\zeta_{++} . \label{zzz}
\end{equation}
Here also, the above bounds obtained on COP are new bounds which are not previously reported elsewhere. Comparing Eqs.
(\ref{bound high wc}) and (\ref{zzz}), we can conclude that the bounds obtained above lie above the parametric region  covered by  the COP curves given in Eq. (\ref{bound high wc}). We also obtain the corresponding expression for the COP
of the SSD refrigerator for the optimization of $\chi$-criterion. It is given by, $\zeta^{\chi}_{++}=\sqrt{4+3\zeta_C}-2$.

\subsubsection{Series behavior of the COPs} 
We further extend our study by analyzing the  near-equilibrium series expansions of various COP expressions obtained at MOF 
(summarized in Table \ref{tab:eng-maaath}, Column I). These series expansions show very interesting behavior which is absent in the series expansions of various corresponding COPs obtained in the optimization of $\chi$ criteria [Table \ref{tab:eng-maaath}, Column III] \cite{VJ2020}. The first term ($2\zeta_C/3$) is same for all the COPs and the second terms form an  arithmetic series with a common difference of  1/18. More interestingly, the differences of  third terms constitute an arithmetic series with a common difference  $1/216 \zeta_C$. To complement our findings, we also report the series expansions  of various forms of efficiencies [Table \ref{tab:eng-maaath}, Column II] obtained  under the similar conditions for the optimization of  $\Omega$ function for the SSD engine\cite{VJ2019}, and observe exactly similar behavior. Although in Ref. \cite{VJ2019}, the authors derived the various forms of efficiencies, they did not analyze the series behavior. 
Further, for the optimization of $\chi$-criterion, we can only say that the leading order term in the  series is proportional 
to $\sqrt{\zeta_C}$ \cite{ShengTu2013}.

\section{\label{sec:level4} Cooling POWER AT MAXIMUM $\Omega$ FUNCTION VERSUS Maximum COOLING POWER}
In this section, we compare the CP obtained at  MOF to the maximum CP. As CP can be optimized 
with respect to $\omega_c$ only \cite{VJ2020}, we can compare this case only. As a representative of our results, 
we confine our discussion to the high-temperature and strong matter-field coupling regime. In this regime, 
the expressions for optimal CP were derived in Ref. \cite{VJ2020}, and are given by:
\begin{equation}
\dot{Q}^*_{c(\gamma\to\infty)}=\frac{2\hbar\Gamma_c \omega_h}{3}\frac{\zeta_C}{1+\zeta_C},
\quad
\dot{Q}^*_{c(\gamma\to0)}=\frac{\hbar\Gamma_h\omega_h}{6}\frac{\zeta_C}{1+\zeta_C}. \label{optimalcp}
\end{equation}
Dividing Eq. (\ref{cpinfinity}) by equation on left hand side of Eq. (\ref{optimalcp}), we obtain the ratio of CP at MOF to the optimal CP,
\begin{equation}
R_{\gamma\to\infty}=1-\sqrt{\frac{1+\zeta_C}{2+\zeta_C}}, \label{ratio1}
\end{equation}
which approaches the value $1-1/\sqrt{2}(=0.29)$ for small values of $\zeta_C$, while it vanishes for large $\zeta_C$.

Similarly, dividing Eq. (\ref{cpzero}) by equation on right hand side of Eq. (\ref{optimalcp}), we obtain the corresponding  ratio for $\gamma\to0$,
\begin{equation}
R_{\gamma\to0}=\frac{3+2\zeta_C}{(2+\zeta_C)^2}, \label{ratio2}
\end{equation}
which approaches the value 3/4 for small $\zeta_C$, while it vanishes for large $\zeta_C$.
We have plotted the Eqs. (\ref{ratio1}) and (\ref{ratio2}) in Fig.3, from which it is clear that ratio is greater for the limiting case $\gamma\to0$. Interestingly, though both ratios vanish as $\zeta_C\to\infty$, their ratio $R_{\gamma\to0}/R_{\gamma\to\infty}$ is finite and approaches to 4 for $\zeta_C\to\infty$. Comparing Eqs. (\ref{ratio1}) and
(\ref{ratio2}),  we can conclude that a relatively large system-bath coupling ($\Gamma_c>>\Gamma_h$ or $\gamma\to0$) yields a
higher relative value of the CP (see Fig. 3).
\begin{figure} 
 \begin{center}
\includegraphics[width=8.6cm]{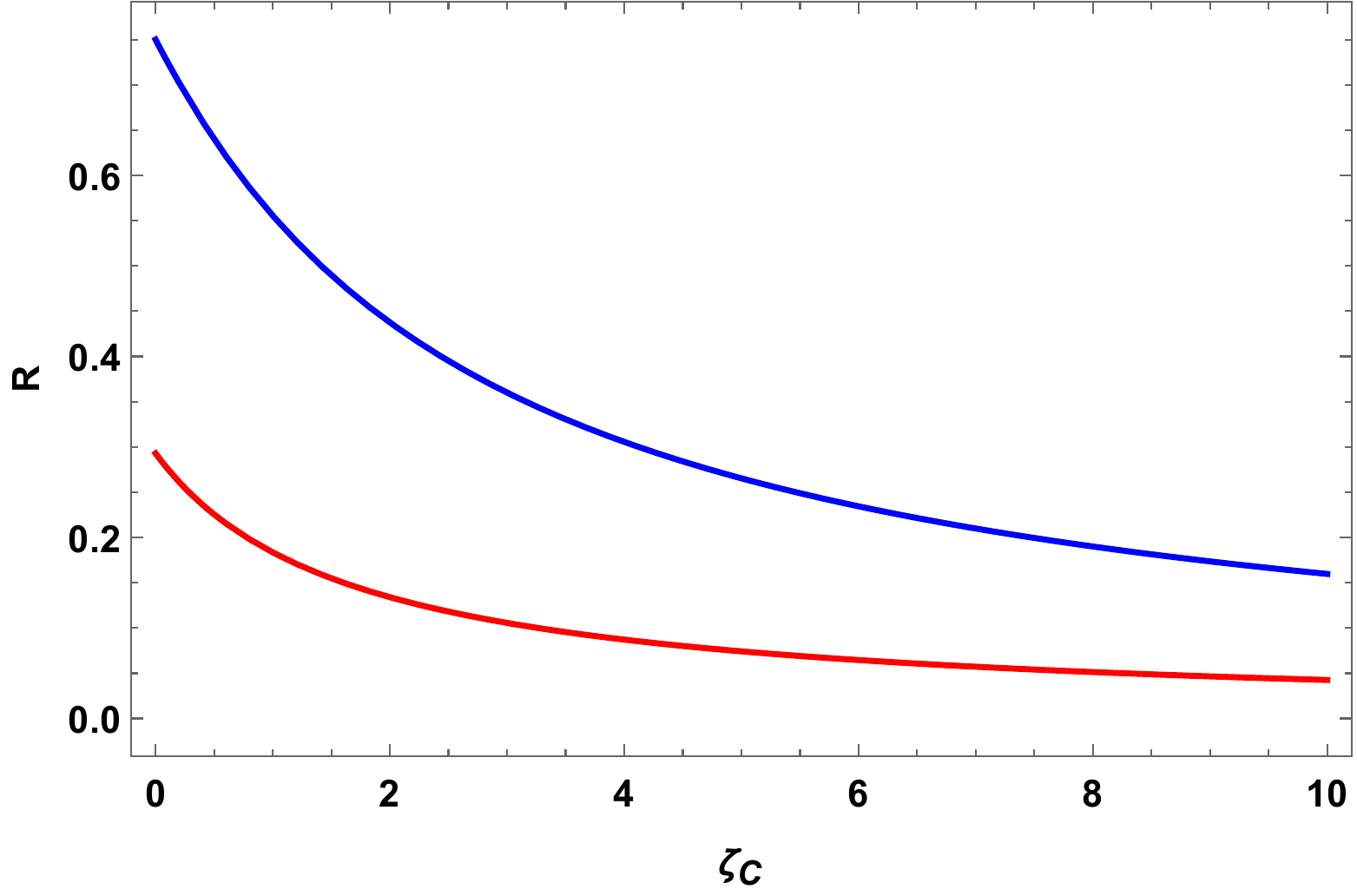}
 \end{center}
\caption{Ratio of the CP at MOF to the maximum CP. Red and blue curves represent
Eqs. (31) and (32), which approach the value $1-1/\sqrt{2}$ and 3/4, respectively for $\zeta_C\to0$, and 
vanish for $\zeta_C\to\infty$.}
\label{final}
\end{figure}

Further, to observe the behavior of the CP at MOF, we plot Eqs. (B8) and (B9) as a 
function of $\zeta_C$ in Fig. 4. 
It is clear from Fig. 4 that the maximum of the CP exists at some value of Carnot COP $\zeta_C$ for both 
limiting cases ($\gamma\to0,\infty$). This suggests that when we operate our refrigerator at MOF, 
temperatures of the reservoirs can always be chosen in such a way that they correspond to the maximum 
CP achievable. In this way, we can choose set the optimal operational point for the thermal device
under consideration.
\begin{figure}  
 \begin{center}
\includegraphics[width=8.6cm]{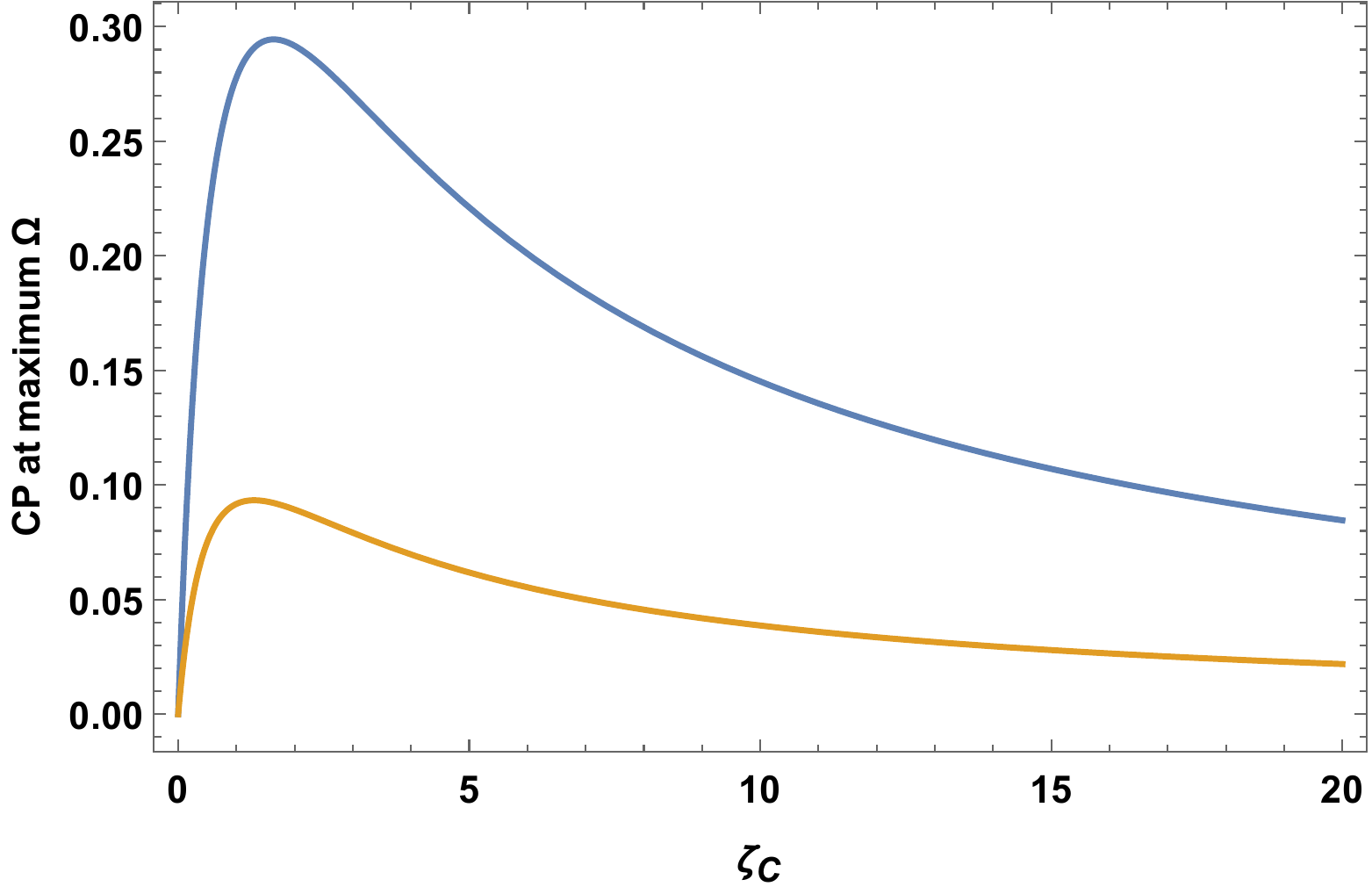}
 \end{center}
\caption{Plot of the scaled CP at MOF as a function of Carnot COP $\zeta_C$. Blue and red
curves represent Eqs. (B8) and (B9), respectively.}
\label{final}
\end{figure}


\section{\label{sec:level5} CONCLUSION}
In this paper, we have analyzed  the performance of a three-level quantum refrigerator operating 
under a unified trade-off objective function known as $\Omega$ function. First, we carried out a 
two-parameter optimization of the $\Omega$ function with respect to the control frequencies 
$\omega_c$ and $\omega_h$) in the low-temperature regime. The obtained form of the COP is independent 
of the parameters of the  system under consideration and depends only on the ratio, $\tau$, of reservoir temperatures.
Interestingly, the first two terms in the series expansion of the COP match exactly with the COPs of 
endoreversible (low-dissipation) and minimally nonlinear irreversible models of refrigerator.
Then, by employing the one-parameter optimization scheme in the high temperature regime, we obtained 
analytic expressions for the lower and upper bounds on the COP in strong as well as weak (intermediate) matter-field coupling conditions. Under the  strong matter-field coupling condition, the obtained bounds 
match with those of classical models of refrigerators. However, for weak (intermediate) matter-field coupling condition,
we obtained new bounds on the COP  which lie beyond the area covered by COP bounds obtained in 
strong matter-field coupling regime. Further, we observed that the first term is same in series expansions
of the various forms of COPs obtained in high temperature regime, which is quite remarkable. Finally, we closed 
our analysis by making a comparison between the SSD refrigerator operating in the maximum CP regime
and the refrigerator operating at MOF.

\section{ACKNOWLEDGEMENTS}
K.K. acknowledges financial support in the form of Postdoctoral Research Fellowship from Indian Institute of Science Education and Research, Mohali
\appendix\section{Steady state solution of density matrix equations}
Here we solve density matrix in steady state. Substituting expressions for $H_0,\bar{H},V_R$ and using Eqn.(5) and (6) in (7) the time evolution of elements of density matrix are governed by following equations
\begin{equation}
\dot{\rho_{11}}=\iota\lambda(\rho_{10}-\rho_{01})-2\Gamma_h[(n_h+1)\rho_{11}-n_h\rho_{gg}],
\end{equation}
\begin{equation}
\dot{\rho_{00}}=-\iota\lambda(\rho_{10}-\rho_{01})-2\Gamma_c[(n_c+1)\rho_{00}-n_c\rho_{gg}],
\end{equation}
\begin{equation}
\dot{\rho_{10}}=-[\Gamma_h(n_h+1)+\Gamma_c(n_c+1)]\rho_{10}+\iota\lambda(\rho_{11}-\rho_{00}),
\end{equation}
\begin{equation}
\rho_{11}=1-\rho_{00}-\rho{gg},
\end{equation}
\begin{equation}
\dot{\rho_{01}}=\dot{\rho_{10}^\star}.
\end{equation}
Solving Eqns.(A1) to (A5) in steady state by setting $\dot{\rho_{mn}}=0(m,n=0,1)$, we obtain
\begin{widetext}
\begin{equation}
\rho_{10} = \frac{i\lambda(n_h-n_c)\Gamma_c\Gamma_h}
{
\lambda^2[(1+3n_h)\Gamma_h + (1+3n_c)\Gamma_c] + \Gamma_c\Gamma_h[1+2n_h+n_c(2+3n_h)][(1+n_c)\Gamma_c + (1+n_h)\Gamma_h ] 
},
\end{equation}
\end{widetext}
and
\begin{equation}
\rho_{01}=\rho_{10}^\star.
\end{equation}
Calculating the trace in Eqns.(10) and (11) the output power and cooling power are written as
\begin{equation}
P=i\hbar\lambda(\omega_h-\omega_c)(\rho_{01}-\rho_{10}),
\end{equation} 
\begin{equation}
\dot{Q_C}=i\hbar\lambda\omega_c(\rho_{10}-\rho_{01}).
\end{equation}
Now $\Omega$ is given by
\begin{equation}
\Omega =2\dot{Q_c}-\zeta_CP.
\end{equation}
Using Eqn.(A8) and (A9) we can write (A10) as
\begin{equation}
\Omega =i\hbar\lambda((2+\zeta_C)\omega_c-\zeta_C\omega_h)(\rho_{10}-\rho_{01}).
\end{equation}
Using Eqns.(A6) and (A7) in this and (A9) we obtain the following expressions for $\Omega$ and CP respectively
\begin{widetext}
\begin{eqnarray}
\Omega =\frac{2\lambda^{2}(n_c-n_h)[(2+\zeta_C)\omega_c-\zeta_C\omega
_h]\Gamma_c\Gamma_h}{\lambda^2[(1+3n_h)\Gamma_h+(1+3n_c)\Gamma_c]+\Gamma_c\Gamma_h[1+2n_h+n_c(2+3n_h)][(1+n_c)\Gamma_c+(1+n_h)\Gamma_h]},\\
\dot{Q_c}=\frac{2\hbar\lambda^2\Gamma_c\Gamma_h(n_c-n_h)\omega_c}{\lambda^2[(1+3n_h)\Gamma_h+(1+3n_c)\Gamma_c]+\Gamma_c\Gamma_h[1+2n_h+n_c(2+3n_h)][(1+n_c)\Gamma_c+(1+n_h)\Gamma_h]}.
\end{eqnarray} 
\end{widetext}

\section{Optimization in the high-temperature regime}
In high-temperature and  strong matter-field coupling regime, we set $n_{h,c} = 1/(e^{\omega_{h,c}/T_{h,c}}-1)
\approx T_{h,c}/\omega_{h,c}\gg 1$. Then Eq. (A12) can be approximated by the following equation
\begin{equation}
\Omega =\frac{2\Gamma_h(\tau\omega_h-\omega_c)((2+\zeta_C)\omega_c-\zeta_C\omega_h)}{3(\gamma\omega_c+\tau\omega_h)},
\label{ax0}
\end{equation}
Setting $\partial \Omega/\partial\omega_c=0$, the optimal solution for $\omega_c$ is obtained as
\begin{equation}
\omega_c^* = \frac{\tau\Big(2-\tau-\sqrt{(1+\gamma)(2-\tau)(2+\gamma-\tau)}  \Big)\omega_h}
{\gamma(\tau-2) }, \label{ax111}
\end{equation}
Substituting Eq. (\ref{ax111}) in Eq. (11), we obtain the following   expression for the COP at MOF: 
\begin{equation}
\zeta_{\omega_h}= \frac{\tau\Big(2-\tau-\sqrt{(1+\gamma)(2-\tau)(2+\gamma-\tau)}  \Big)}
{\gamma(\tau-2)-\tau\Big(2-\tau-\sqrt{(1+\gamma)(2-\tau)(2+\gamma-\tau)}  \Big)}, \label{ax1}
\end{equation}
%
%
Now, we will make use of the limiting forms of $\omega_c^*$ which can be obtained by taking the limits 
$\gamma\to\infty$ and $\gamma\to0$ in Eq. (\ref{ax111}), and are given by following equations, respectively:
\vspace{3mm}
\begin{eqnarray}
\omega_c^*(\infty) &=& \frac{\zeta_C}{(1+\zeta_C)(2+\zeta_C)}\omega_h, \label{b4}
\\
\omega_c^*(0) &=& \frac{\zeta_C(3+2\zeta_C)}{4(1+\zeta_C)(2+\zeta_C)^2}\omega_h, \label{b5}
\end{eqnarray}
Further, using Eqs. (\ref{b4}) and (\ref{b5}) in Eq. (\ref{ax0}), we obtain following expressions
for the optimal $\Omega$ function for the limiting cases $\gamma\to\infty$ and $\gamma\to0$, respectively: 
\begin{equation}
\Omega^*_{\gamma\to\infty} =  \frac{2\hbar\Gamma_c}{3} \left( \frac{\zeta_C\big(3+2\zeta_C-2\sqrt{(1+\zeta_C)(2+\zeta_C)}\big)}{(1+\zeta_C)} \right)\omega_h,
\end{equation}
\begin{equation}
\Omega^*_{\gamma\to0} =\frac{ \hbar\Gamma_h}{6} \left( \frac{\zeta_C }
{ (1+\zeta_C)(2+\zeta_C)}\right)\omega_h.
\end{equation}
The corresponding expressions for the cooling power, $\dot{Q}_c=2\hbar\Gamma_h(\tau\omega_h-\omega_c)/3(\gamma\omega_c+\tau\omega_h)$ , at optimal $\Omega$ function are given by:
\begin{equation}
\dot{Q}^\Omega_{c(\gamma\to\infty)} = \frac{2\hbar\Gamma_c}{3}
\left(\frac{\zeta_C}{1+\zeta_C}-\frac{\zeta_C}{\sqrt{(1+\zeta_C)(2+\zeta_C)}}\right)\omega_h, \label{cpinfinity}
\end{equation}
\begin{equation}
\dot{Q}^\Omega_{c(\gamma\to0 )} =\frac{ \hbar\Gamma_h}{6} \left(\frac{\zeta_C(3+2\zeta_C)} 
{ (1+\zeta_C)(2+\zeta_C)^2}\right)\omega_h.\label{cpzero}
\end{equation}
\section{Two parameter optimization in the low-temperature limit}
Using Eq. (\ref{omega low}), setting $\partial\Omega/\partial\omega_c = 0$ and $\partial\Omega/\partial\omega_h = 0$, we get the following set of equations:

\begin{eqnarray}
\omega_c - \frac{\zeta_C}{2+\zeta_C}\omega_h &=& \left(1 - e^{\omega_{c}/T_{c}-\omega_{h}/T_{h}}\right)T_c  \label{C1}
\\
\omega_c - \frac{\zeta_C}{2+\zeta_C}\omega_h &=& \frac{\left(e^{-\omega_{c}/T_{c}+\omega_{h}/T_{h}}-1\right)\zeta_C T_h}{2+\zeta_C} \label{C2}
\end{eqnarray}
On comparing Eqs. (\ref{C1}) and (\ref{C2}), we get
\begin{equation}
e^{\frac{\omega_c}{T_c}- \frac{\omega_h}{T_h}}   =   \frac{1+\zeta_C}{2+\zeta_C}
\end{equation}
Taking log on both sides, we have
\begin{equation}
\frac{\omega_c}{T_c} - \frac{\omega_h}{T_h}   =   \ln \left(\frac{1+\zeta_C}{2+\zeta_C}\right)
\label{C3}
\end{equation}
Substitute Eq. (\ref{C3}) in Eq. (\ref{C1}), we get
\begin{equation}
\frac{\omega_c}{T_c} - \frac{1+\zeta_C}{2+\zeta_C}\frac{\omega_h}{T_h}   =  \frac{1}{2+\zeta_C}
\label{C4}
\end{equation}
Now, solving Eqs. (\ref{C3}) and (\ref{C4}) simultaneously, we obtain the optimal expressions for $\omega_c$ and $\omega_h$ given by Eq. (\ref{wc wh}).

\bibliography{biblo1}
\bibliographystyle{apsrev4-1}

\end{document}